\documentclass[icps3]{svjour}
\usepackage{graphicx}
\begin{document}
\title{Critical Behavior of the Thermoelectric Transport Properties
in Amorphous Systems near the Metal-Insulator Transition}
\titlerunning{Critical Behavior of the Thermoelectric Transport Properties}
\author{C. Villagonzalo\inst{1} \thanks{ \emph{Permanent address:
      National Institute of Physics, University of the Philippines,
      Diliman, 1101 Q. C., Philippines} }
          \and R. A.  R\"{o}mer\inst{1} \and M. Schreiber\inst{1} \and A.
      MacKinnon\inst{2}
}                     % Do not remove
\authorrunning{C. Villagonzalo, et al.}
\institute{Institut f\"{u}r Physik, Technische Universit\"{a}t,
 09107 Chemnitz, Germany
 \and Blackett Laboratory, Imperial College, Prince Consort Rd., London 
SW7 2BZ, U.K.}
\maketitle
\begin{abstract}
  The scaling behavior of the thermoelectric transport properties in
  disordered systems is studied in the energy region near the
  metal-insulator transition. Using an energy-dependent conductivity
  $\sigma$ obtained experimentally, we extend our
  linear-response-based transport calculations in the
  three-dimensional Anderson model of localization. Taking a dynamical
  scaling exponent $z$ in agreement with predictions from scaling
  theories, we show that the temperature-dependent $\sigma$, the
  thermoelectric power $S$, the thermal conductivity $K$ and the
  Lorenz number $L_0$ obey scaling.
\end{abstract}
%
%%%%%%%%%%%%%%%%%%%%%%%%%%%%%%%%%%%%%%%%%%%%%%%%%%%%%%%%%%%%%%%%%%%%%%%%%%%%%%
\vspace{0.4cm}

%%% Introduction

\noindent The scaling description \cite{AbrALR79} of
disordered systems, e.g.\ the Anderson model of localization, has
cultivated our understanding of transport properties in such systems
\cite{KraM93,LeeR85}.  According to the scaling hypothesis, the
behavior of the d.c.\ conductivity $\sigma$ near the metal-insulator
transition (MIT) in the Anderson model can be described by only a
single scaling variable. As a result of the scaling theory, the
dynamical conductivity in the three-dimensional (3D) Anderson model
behaves as \cite{Weg76,BelK94}
\begin{equation}
\frac{\sigma(t,T)}{T^{1/z}}={\mathcal{F}} 
\left(\frac{t}{T^{1/\nu z}}\right)\;, \label{dc_scaling} 
\end{equation} 
where $T$ is the temperature and $t$ is the dimensionless distance
from the critical point. For example, $t = \left|1-E_F/E_c\right|$
where $E_F$ and $E_c$ are the Fermi energy and the mobility edge,
respectively.  The parameter $\nu$ is the correlation length exponent,
which in 3D is equivalent to the conductivity exponent, $\sigma
\propto t^{\nu}$, and $z$ is the dynamical exponent, $\sigma\propto
T^{1/z}$. It was further demonstrated that not only $\sigma(t,T)$
obeys scaling in the 3D Anderson model but also the thermoelectric
power $S(t,T)$ \cite{SivI86,VilRS99b}, the thermal conductivity
$K(t,T)$ and the Lorenz number $L_0(t,T)$ \cite{VilRS99b}.  
However, despite the quality of the scaling of $\sigma$, we obtained
an unphysical value for $z$ \cite{VilRS99b}. Scaling arguments for
noninteracting systems predict $z=d$ in $d$ dimensions
\cite{Weg76,BelK94}. But we found \cite{VilRS99b} $z=1/\nu\ll 3$.  In
addition, values of $S(T)$ \cite{EndB94,VilRS99a} are at least an
order of magnitude larger than in measurements of doped semiconductors
\cite{LakL93} and amorphous alloys \cite{LauB95,SheHM91}.

In what follows, we show that we obtain the right order of magnitude
\cite{VilRSM00} and good scaling for these thermoelectric transport
properties by using a "modified" critical behavior of $\sigma$ in the
linear-response formulation for the Anderson model based on
experimental data.

%%% Theoretical Background

In the linear-response formulation, the thermoelectric transport
properties can be determined from the kinetic coefficients $L_{ij}$
\cite{VilRS99a}, i.e.,
\begin{eqnarray}
\sigma = L_{11}, \hspace{8mm} & &
K=\frac{L_{22}L_{11}-L_{21}L_{12}}{e^{2}TL_{11}} \;,
 \nonumber\\  \label{lreq} \\ 
S = \frac{L_{12}}{|e|TL_{11}}\;, & \hspace{3mm}\mbox{and}\hspace{3mm} &
L_{0}=\frac{L_{22}L_{11}-L_{21}L_{12}}{(k_{B}TL_{11})^2}\,. \nonumber 
\label{eq-tetp}
\end{eqnarray}
The $L_{ij}$ relate the induced charge and heat current densities
to their sources such as a temperature gradient \cite{VilRS99a}. 
In the absence of interactions and inelastic scattering
processes, the $L_{ij}$ are expressed as \cite{CheT61,Kub57,Gre58}
\begin{eqnarray} 
L_{ij}= (-1)^{i+j}\int_{-\infty}^{\infty} A(E)
\left[E-\mu(T)\right]^{i+j-2}  \nonumber \\
\cdot \left[-\frac{\partial f(E,\mu,T)}{\partial E} \right] dE\;, 
\label{coeff}
\end{eqnarray} 
for $i,j=1,2$, where $\mu$ is the chemical potential of the system,
$f(E,\mu,T)$ is the Fermi distribution function, and $A(E)$ describes
the system dependent features. In the Anderson model, one sets $A(E)$
to be equal to the critical behavior of $\sigma(E)\propto
|1-E/E_c|^{\nu}$ \cite{KraM93}.  Note, however, that this behavior
near the MIT does not contain a $T$ dependence. Hence, the $T$
dependence of the $L_{ij}$ in Eq.\ (\ref{coeff}) is merely due to the
broadening of $f$ and the $T$ dependence of $\mu$.  The latter stems
from the structure of the density of states, variations in which yield
only negligible changes in $L_{ij}$ \cite{VilRSM00}. Thus, in order to
model a correct $T$ dependence of the $L_{ij}$ and, consequently, the
transport properties, we need to reconsider what $A(E)$ should be.

%%% Method

A suitable $\sigma(E)$ may be obtained from appropriate experimental
data.  The recent measurements of $\sigma$ by Waffenschmidt \textit{et
  al}.\ \cite{WafPL99} in Si:P at the MIT under uniaxial stress show
that $\sigma(t,T)$ obeys scaling with $\nu=1 \pm 0.1$ and $z=2.94 \pm
0.3$.  Obtaining (i) $z \approx d = 3$ in good agreement with scaling
arguments \cite{Weg76,BelK94} and (ii) $\nu$ which also agrees
reasonably well with the numerical results for noninteracting systems
\cite{SleO99,MilRS99a,MilRSU00} makes the experimental data in Ref.\ 
\cite{WafPL99} an excellent empirical model for $A(E)$. In those
experiments, $t$ in $\sigma(t,T)$ is given in terms of the uniaxial
stress and its critical value near the MIT.  Here, we derive a
functional form of $\sigma(E)$ by constructing a polynomial fit or a
spline curve of the experimental data and setting $t=|1-E/E_c|$. When
using this data as input for Eq.\ (\ref{eq-tetp}) the difference in
the order of magnitude in $S$ as compared to experiments is removed
\cite{VilRSM00}. Thus, following the approach of Ref.\ 
\cite{VilRSM00}, we now study whether the $T$ dependence of
$\sigma(E,T)$ can be scaled as in Eq.\ (\ref{dc_scaling}) and whether
$S$, $K$ and $L_0$ also obey scaling.

%%% Results

%%%%%%%%%%%%%%%%%%%%%%%%%%%%%%% FIGURE %%%%%%%%%%%%%%%%%%%%%%%%%%%%%%%%%%%%%%
\begin{figure}  
\includegraphics[width=\hsize]{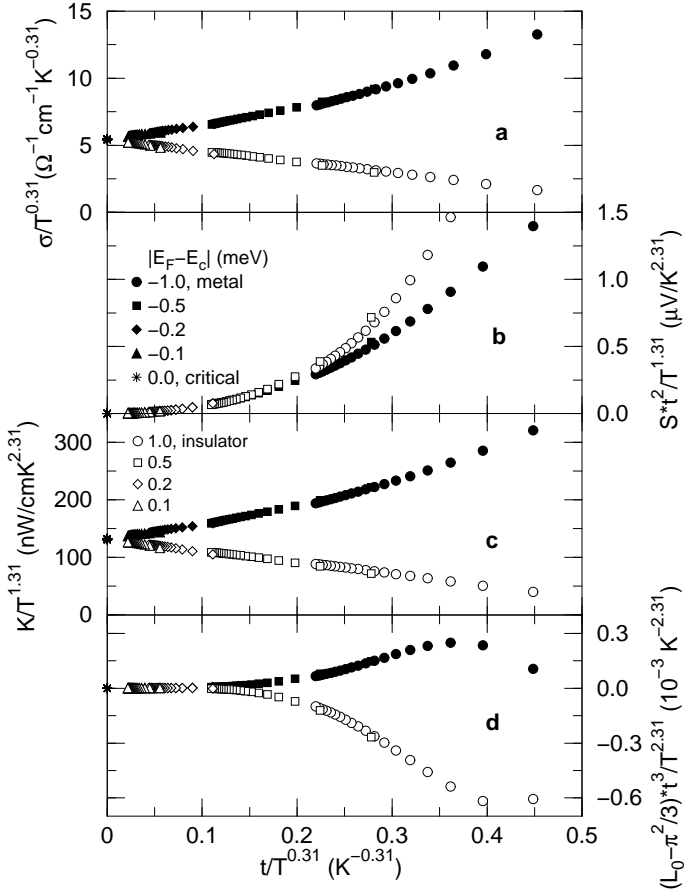}
\caption{Scaling of thermoelectric transport properties
where $t=|1-E_F/E_c|$. The different symbols denote
the relative positions of various values of the Fermi energy $E_F$ 
with respect to the mobility edge $E_c$.}
\label{fig:1}       
\end{figure}  
%%%%%%%%%%%%%%%%%%%%%%%%%%%%%%% END OF FIGURE %%%%%%%%%%%%%%%%%%%%%%%%%%%%%%%

In Fig.\ \ref{fig:1} we show that $\sigma$, $S$, $K$ and $L_0$ data
for different $t$ and $T$ parameters collapse onto scaling curves when
plotted as a function of $|1-E_F/E_c|/T^{0.31}$. For each figure, we
clearly obtain two branches, one for the metallic regime and another
for the insulating regime. As depicted in Fig.\ \ref{fig:1}a, $\sigma$
satisfies Eq.\ (\ref{dc_scaling}). With $\nu=1$ in accordance with the
experiment in Ref.\ \cite{WafPL99}, $1/\nu z = 0.31$ gives $z=3.2$.
This is in good agreement with the prediction $z=3$ for 3D
noninteracting systems. Furthermore, it indicates that the Harris
criterion \cite{Har74}, $\nu z > 1$, is satisfied which in turn
implies a sharp MIT.

The prefactor in $K$ in  Fig.\ \ref{fig:1}c verifies that $\sigma$
and $K/T$ behave similarly as the MIT is approached. This confirms
earlier results in the 3D Anderson model from various methods
\cite{VilRS99a,StrC87}. Meanwhile, the prefactors in both $S$ and
$L_0$ in Figs.\ \ref{fig:1}b and \ref{fig:1}d have not been observed
in the respective scaling curves in the 3D Anderson model
\cite{VilRS99b}. The prefactors serve as corrections to the scaling of
$S$ and $L_0$ when an appropriate $\sigma(E,T)$ is used as input in 
Eq.\ (\ref{coeff}).

As shown in Fig.\ \ref{fig:1}d in accordance also with the results in
Ref.\ \cite{VilRSM00}, $L_0\rightarrow \pi^2/3$, as the MIT is
approached from the metallic or the insulating regime.  This is the
expected value in the Sommerfeld free electron theory. It is different
from the result for the unmodified 3D Anderson model \cite{VilRS99a}
for which the magnitude of $L_0$ depends on $\nu$ \cite{VilRS99a}.

%%% Conclusion

%In conclusion, we find that by $\sigma(E,T)$ in our calculations, the
%behavior of the thermoelectric transport properties near the MIT obey
%scaling.

In conclusion, we find that by modifying $\sigma(E,T)$ in our
calculations, the thermoelectric transport properties
near the MIT obey scaling.

%%%%%%%%%%%%%%%%%%%%%%%%%%%%%%%%%%%%%%%%%%%%%%%%%%%%%%%%%%%%%%%%%%%%%%%%%%%%%%
\begin{acknowledgement}
  The authors are grateful for the support of the DFG through
  Sonderforschungsbereich 393, the DAAD, the British Council and the
  SMWK.
\end{acknowledgement}

%\bibliographystyle{prsty}
%\bibliography{bibliograph}

\begin{thebibliography}{99}

\bibitem{AbrALR79}
E. Abrahams, P.~W. Anderson, D.~C. Licciardello, and T.~V. Ramakrishnan, Phys.
  Rev. Lett. {\bf 42}, (1979) 673.

\bibitem{KraM93}
B. Kramer and A. MacKinnon, Rep. Prog. Phys. {\bf 56},  (1993) 1469.

\bibitem{LeeR85}
P.~A. Lee and T.~V. Ramakrishnan, Rev. Mod. Phys. {\bf 57}, (1985)  287.

\bibitem{Weg76}
F. Wegner, Z. Phys. B {\bf 25}, (1976) 327.

\bibitem{BelK94}
D. Belitz and T.~R. Kirkpatrick, Rev. Mod. Phys. {\bf 66}, (1994) 261.

\bibitem{SivI86}
U. Sivan and Y. Imry, Phys. Rev. B {\bf 33},  551  (1986).

\bibitem{VilRS99b}
C. Villagonzalo, R.~A. {R\"{o}mer}, and M. Schreiber, Ann. Phys. (Leipzig) {\bf
  8}, (1999)  SI-269.

\bibitem{EndB94}
J.~E. Enderby and A.~C. Barnes, Phys. Rev. B {\bf 49}, (1994) 5062.

\bibitem{VilRS99a}
C. Villagonzalo, R.~A. {R\"{o}mer}, and M. Schreiber, Eur. Phys. J. B {\bf 12},
   (1999) 179.

\bibitem{LakL93}
M. Lakner and H. v.~{L\"{o}hneysen}, Phys. Rev. Lett. {\bf 70}, (1993) 3475.

\bibitem{LauB95}
C. Lauinger and F. Baumann, J. Phys.: Condens. Matter {\bf 7}, (1995) 1305.

\bibitem{SheHM91}
G. Sherwood, M.~A. Howson, and G.~J. Morgan, J. Phys.: Condens. Matter {\bf 3},
  (1991) 9395.

\bibitem{VilRSM00} C. Villagonzalo, R.~A. {R\"{o}mer}, M. Schreiber 
and A. MacKinnon, preprint (2000), cond-mat/0006083.

\bibitem{CheT61}
G.~V. Chester and A. Thellung, Proc. Phys. Soc. {\bf 77}, (1961) 1005.

\bibitem{Kub57}
R. Kubo, J. Phys. Soc. Japan {\bf 12},  (1957) 570.

\bibitem{Gre58}
D.~A. Greenwood, Proc. Phys. Soc. {\bf 71}, (1958) 585.

\bibitem{WafPL99}
S. Waffenschmidt, C. Pfleiderer, and H. v.~{L\"{o}hneysen}, Phys. Rev. Lett.
  {\bf 83},  (1999) 3005.

\bibitem{SleO99} K. Slevin and T. Ohtsuki, Phys. Rev. Lett. {\bf 82},
  (1999) 382; {\bf 82}, (1999) 669.

\bibitem{MilRS99a}
F. Milde, R.~A. {R\"{o}mer}, and M. Schreiber, Phys. Rev. B {\bf 61}, 
  (2000) 6028.

\bibitem{MilRSU00}
F. Milde, R.~A. {R\"{o}mer}, M. Schreiber, and V. Uski, Eur. Phys. J. B,  
  (2000), 685.

\bibitem{Har74}
A.~B. Harris, J. Phys. C \textbf{7}, (1974), 1671.

\bibitem{StrC87}
G. Strinati and C. Castellani, Phys. Rev. B \textbf{36}, (1987) 2270.

\end{thebibliography}

\end{document}